\documentclass{emulateapj}

\begin{document}

\title{Evolution of reconnection along an arcade of magnetic loops}

\author{Paolo C. Grigis and Arnold O. Benz}
\affil{Institute of Astronomy, ETH Z\"urich, 8092 Z\"urich, Switzerland}
\email{pgrigis@astro.phys.ethz.ch}

\begin{abstract}

RHESSI observations of a solar flare showing continuous motions of double
hard X-ray sources interpreted as footpoints of magnetic loops are presented.
The temporal evolution shows many distinct emission peaks of duration of some
tens of seconds ('elementary flare bursts').
Elementary flare bursts have been interpreted as instabilities or oscillations
of the reconnection process leading to an unsteady release of magnetic energy.
These interpretations based on two-dimensional concepts cannot explain these
observations, showing that the flare elements are displaced in a third
dimension along the arcade. Therefore, the observed flare elements are not
a modulation of the reconnection process, but originate as this process
progresses along an arcade of magnetic loops. Contrary to previous reports,
we find no correlation between footpoint motion and hard X-ray flux.
This flare apparently contradicts the predictions of the standard
translation invariant 2.5D reconnection models.
\end{abstract}


\section{Introduction}
High-energetic electrons accelerated during solar flares emit
bremsstrahlung hard X-rays (HXR), whose evolution in time can
be followed in images, light curves, and spectra. This information is combined
here to infer characteristics of the unknown acceleration process of these particles.

As for light curves, quasi-periodic modulations of the HXR flux on
time scales of some tens of seconds (\emph{Elementary Flare Bursts}; EFB)
have long been known to observers \citep{Parks69,deJager78}.
Spectral studies show that EFBs preferentially follow a characteristic
\emph{soft-hard-soft} spectral behavior
\citep[][and references therein]{Grigis04}.
The modulations of the HXR and microwave flux have been interpreted
in 2D models of reconnection as fluctuations in the reconnection process due
to global oscillations of the loop \citep{Roberts84,Nakariakov03,Stepanov04}.

On the imaging front, early observational evidence from the
\emph{Solar Maximum Mission} Hard X-Ray Imaging Spectrometer
\citep{Hoyng81} showed that HXR sources are found at the location of
footpoints (FP) of magnetic loops.
Different types of HXR sources located higher in the solar
atmosphere were detected on some occasions:
the \emph{Yohkoh} Hard X-ray Telescope registered
fainter \emph{loop-top} sources above the soft X-ray loops
\citep{Masuda94} and the \emph{Reuven Ramaty High Energy Solar Spectrometric
Imager} \citep[RHESSI; ][]{Lin02} observed \emph{coronal} sources
\citep{Lin03}, as well as emission from collisionally thick coronal loops
\citep{Veronig04}. However, the bulk of non-thermal HXRs comes from loop FPs.

Two FPs on opposite sides of a magnetic neutral line are expected in the
standard model of eruptive flares 
\citep[reviewed e.g. by][]{Priest02}: The rapid eruption of a
filament enables the magnetic field to reconnect, driving particle
acceleration in lower loops. Electrons precipitating
to a FP emit HXRs. In this scenario, one expects the observed
FP sources to drift apart as successive field lines
are reconnected at higher altitudes. This explanation fits the long-known outward
motion of $\mathrm{H}\alpha$ ribbons parallel to the neutral line.

Reports on both the morphology and the time evolution of the FPs
show a large range of behaviors: single \citep{Takakura83},
double \citep{Hoyng81} and multiple sources are seen, and many
authors observe FP motions of different kinds: decrease and increase
in the FP separation across the neutral line; parallel and antiparallel
movements along the arcade
\citep[for some recent observations see][]{Fletcher02,Liu04,Qiu04,Siarkowski04}.
This bewildering behavior demonstrates just how complex the flare phenomenon can be.

\cite{Krucker03} presented high-resolution observations of a particularly interesting
two-ribbon flare. One of the HXR FPs moved continuously along a ribbon, whereas the
other two FPs showed no systematic motion. No motion perpendicular to the ribbons are
noticeable, but the parallel motion correlated with the HXR flux. The observed behavior 
allowed Krucker et al. to interpret the observations still in terms of the standard
reconnection model, where the motion is due to receding FPs. This requires a strongly sheared
arcade and a not-specified complex magnetic structure including
the 2 other FPs without systematic motion.

Here HXR source motions observed with RHESSI during
a flare are reported that do not allow for such interpretation by
the standard 2D reconnection model.
We study also the relation between the spatial motion and the
spectral evolution of EFBs in time.

\section{Observations}

\begin{figure}
\plotone{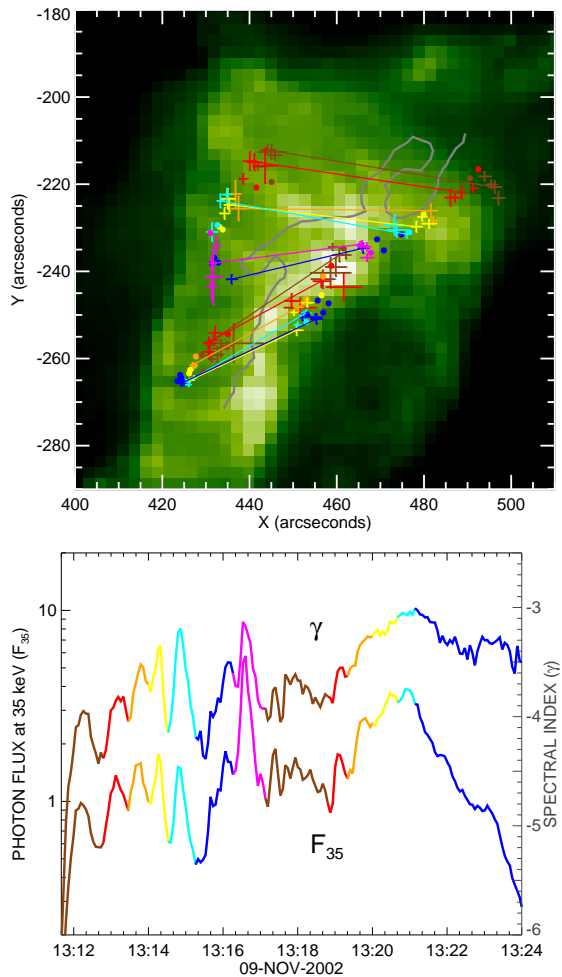}
\caption{\label{fig1}
\emph{Top:} SOHO/EIT 195 {\AA} image of post-flare loops 
with the RHESSI HXR source positions superimposed. The positions of the 20 - 50 keV
sources from the CLEAN images are represented by crosses with arm lengths
proportional to the errors, positions from the PIXON images are given by
circles. Simultaneous footpoints are connected and color coded according
to the time intervals defined in the bottom part. The neutral line is
shown in gray. 
\emph{Bottom:} Time evolution of the flux and spectral index.}
\end{figure}

RHESSI observed the Sun on November 9, 2002 from 12:23 to
13:28 UT, when it entered the shadow of the Earth,
and it registered the HXR evolution of a solar flare
of soft X-ray (GOES) importance M4.9. RHESSI was in a configuration
well suited to the derivation of high-resolution HXR spectra and images:
No decimation of the data was active during the flare and 
attenuation \citep{Smith02} was constantly in state 1
(thick attenuator in), thus ensuring that the detector dead time
was below about 5\% during the flare. Auxiliary data suggest that
this flare was an eruptive event, displaying a post-flare loop arcade
in the SOHO/EIT 195 \AA\ images (Fig.~\ref{fig1}), a moving Type IV radio burst
(Phoenix spectrometer), and an associated fast CME
\citep[listed in the catalog by][]{Yashiro04}.

\begin{figure}
\plotone{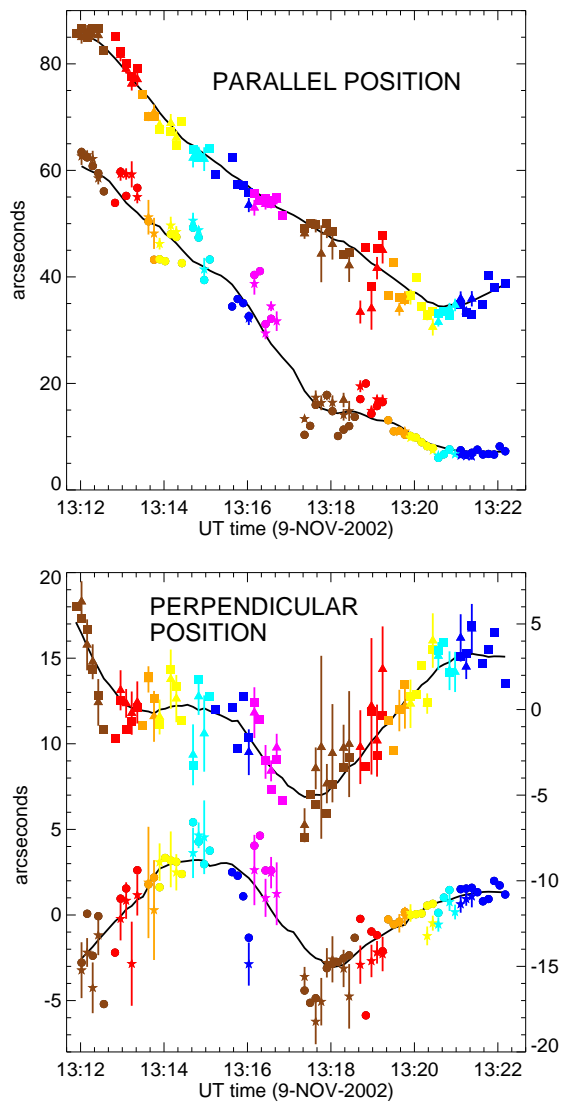}
\caption{\label{fig2}
Time evolution of the source positions relative
to the trend lines. The color code is the same as in Fig.~\ref{fig1},
referring to the major subpeaks. Triangles and stars with error bars refer to values
derived using CLEAN, squares and circles using PIXON, for the western
and eastern FPs, respectively.
\emph{Top:} The upper curve displays the parallel coordinates of the western FPs,
the lower curve the same of the eastern FPs.
\emph{Bottom:} Time evolution of the coordinate perpendicular
to the regression lines. The upper curve refers to the western FP (scale on the right),
the lower curve to the eastern FP (scale on the left).
Both panels show in black the averaged smoothed motion for each FP (PIXON value),
defining a new reference for detailed analysis.}
\end{figure}

Spatially integrated HXR spectra for this event were obtained
at a cadence of one RHESSI rotation period ($\sim 4$ s)
and used in a previous study \citep{Grigis04} to analyze the
time evolution of the non-thermal part. Here we additionally
produce HXR images averaged over two rotation periods ($\sim 8$ s)
in the energy band 20--50~keV using the CLEAN \citep{Hurford02}
and the PIXON \citep{Metcalf96} reconstruction algorithms.
The images resulting from the two different methods were inspected
and compared. We discarded images of poor quality, obtaining
43 CLEAN and 69 PIXON images. Most images show two sources
located at opposite sides of the magnetic neutral line, for some others only
one source is clearly defined. We computed the source
positions in each image by fitting a two-dimensional elliptical
Gaussian to each visible source separately. For the CLEAN images
we were able to estimate the statistical positional error by
dividing the $1\sigma$ source width (provided by the Gaussian fit)
by the signal to noise ratio (obtained dividing the peak flux
by the standard deviation of the fluctuations in the image outside
the sources).
The average error estimated in this way amounts to 1.4$\arcsec$.
This method cannot be applied for the PIXON images, since
the PIXON algorithm suppresses the noise in the image.

The evolution of the positions of the eastern and western
FPs are shown in the top panel of Fig.~\ref{fig1}
superimposed on a SOHO/EIT \citep{Delaboudiniere95} image taken at 13:48 UT
showing a post-flare loop arcade. The crosses represent the
CLEAN positions with their error bars, and the circles the PIXON
positions. We compensate for the effect of the solar rotation by
rotating each source to the position it would assume
at the time of the EIT image.

Both FPs start from the northern part of the image and move
along the two ribbons visible to EIT in the south-east direction.
The northern part of the arcade is wider than its southern end,
and therefore the north-south movement along it effectively
causes a convergence of the opposite FPs.

In the bottom panel of Fig.~\ref{fig1} the time
evolution of the non-thermal HXR flux at 35~keV and the spectral
index are presented. Emission peaks with a duration ranging from a few tens of seconds
to the observational limit at 8 s can be noted, each showing soft-hard-soft behavior.
The main peaks are drawn in a color different from their neighbors such that the source
positions in the top panel, having the same code, can be followed in their temporal evolution.

To characterize the motion along the arcade, we define
an eastern and a western regression line obtained by two
independent least-squares fittings of all the positions of the eastern and western FPs.
The two straight lines go from SE to NW and are not shown in Fig.~\ref{fig1}. The lines
are inclined by $74\degr$ (eastern) and $36\degr$ (western) to the E--W
direction. From now on, every decomposition of a vector in its \emph{parallel}
and \emph{perpendicular} components will refer to the directions
given by the regression lines. The parallel coordinate increases from an arbitrary origin towards
NW, whereas the perpendicular coordinate is positive in the direction
which points away from the arcade.

The motions parallel and perpendicular to the two regression lines are
presented in of Fig.~\ref{fig2}. The FPs move predominantly along the lines,
thus parallel to the ribbons. The parallel motion is quite smooth and continuous,
especially for the western FP. For both FPs, the speed diminishes after
about 13:17 UT. The only large discontinuity in the parallel motion
is a possible $20 \arcsec$ jump of the parallel eastern FP position after the strongest
HXR peak when the eastern source is not detectable. Afterwards, the eastern FP
move slower and get stationary after 13:20 UT. Contrary to previous reports, we find
no correlation between FP speed and HXR flux.

Do subpeaks show motions perpendicularly outward from the ribbons as expected from the standard
reconnection model? In Fig.~\ref{fig2} (bottom) this is not obvious, although the two FPs are
apparently moving relative to the regression line. Note however, that the lines are converging, thus
the effective FP separation decreases. Moreover, the ribbons are not straight. To study the question
in more detail, we additionally define two smooth trend curves following the FP motions more closely.
A moving average of the PIXON positions of each FP branch was computed, using a boxcar smoothing
window of 15 elements, interpolating the missing points. The interval corresponds to a duration
of 120 seconds, longer than all impulsive subpeaks in the HXR light curve (Fig.~\ref{fig1}, bottom).
The smooth trends are shown in Fig.~\ref{fig2} as black continuous curves.

The standard reconnection model predicts outward FP motion at a given place in the arcade.
In order to look for such systematic trends within HXR subpeaks, we took the parallel
and perpendicular components of the difference vector from the smoothed source position
to the observed PIXON positions.
For each subpeak, we averaged the positions occurring during the first half and 
the second half, and calculated the difference second minus first half,
$\Delta_\mathrm{POS}$, for both eastern and western sources.

\begin{figure}
\plotone{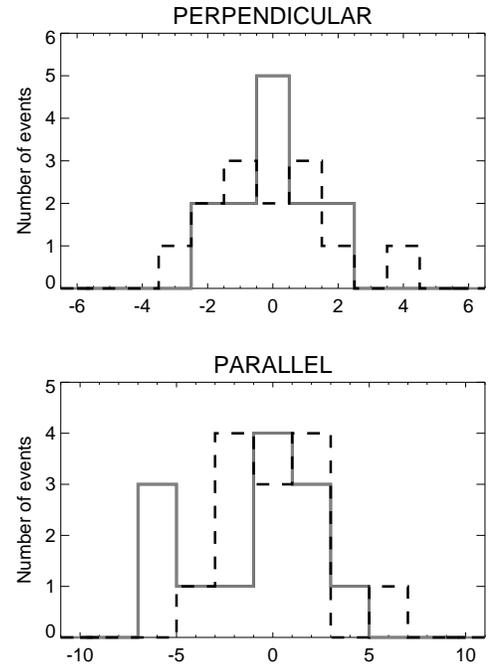}
\caption{\label{fig3}
Distribution of the average motions during a peak in 
perpendicular and parallel directions relative to the time averaged trend curves.
Eastern FPs are shown with continuous lines, western FPs with dashed lines.}
\end{figure}

For elementary flare bursts produced by standard reconnection, one would expect outward
moving sources, thus $\Delta^\perp_\mathrm{POS}$ being positive, at least on the average.
Furthermore, the motion along the ribbons should be stepwise and discontinuous with
$\Delta^\parallel_\mathrm{POS}$ being positive if each Elementary Flare Burst were a
localized event. Figure \ref{fig3}
demonstrates that these expectations are not satisfied during subpeaks of this flare.
The distribution of the average perpendicular motion during each peak shown in
Fig.~\ref{fig3} has a mean $\Delta^\perp_\mathrm{POS}$ value of $0.0\arcsec \pm 0.4 \arcsec$
for the eastern FP and $0.2\arcsec \pm 0.5 \arcsec$ for the western FP (the error is
the standard error of the average). The mean value of the relative parallel motion
during the peaks is  $-1.0\arcsec \pm 1.0 \arcsec$ for the eastern FP and
$0.4 \pm 0.7 \arcsec$ for the western FP.

The global motion along the arcade progresses with an average
velocity in the parallel direction of 63 km s$^{-1}$ for the
eastern FP and 55 km s$^{-1}$ for the western FP. The lower velocity of the
western FP is due to the fact that the latest data points have negative
parallel velocities since they move backwards (Fig.~\ref{fig2}).
Averaging the absolute values of the parallel component of the velocity, we get
65 km s$^{-1}$ for the western FP.
A speed of about 110 km s$^{-1}$ is maintained for 2 minutes in the western FP
at the beginning of the flare, while the data gap and possible jump around 13:17
in the eastern FP position requires 180 km s$^{-1}$.

The line connecting the two FPs is inclined with respect to the post-flare loops
seen in the EIT image (Fig.~\ref{fig1}).
The angle is in the range of $25\degr$ to $70\degr$,
the post-flare loops being nearly perpendicular to the neutral line. This indicates that
the HXR emitting loops are strongly sheared. RHESSI images at lower energy where thermal
emission dominates the spectrum show sources or loops between the FPs. They appear to be
coronal sources moving along the arcade with the FPs at higher energies.

\section{Conclusion}

We have analyzed RHESSI HXR observations of the time evolution
of both images and spectra for a solar flare of GOES class
M4.9. Surprisingly, the footpoints move smoothly along the two ribbons
in contrast to the bursty evolution of the HXR flux. The observed
Elementary Flare Bursts have durations between 30 s and less than 8 s,
and show pronounced spectral soft-hard-soft behavior.
The parallel source motions exclude the generally held notion of
Elementary Flare Bursts being the modulation of a global reconnection process.
Instead, the temporal modulation of the HXR flux and spectral index appear
to be caused by a spatial displacement along the arcade.
This could be caused by some disturbance propagating smoothly along the arcade,
sequentially triggering a reconnection process in successive
loops of the arcade. The disturbance would have to propagate
with a speed in the range 50--150 km s$^{-1}$, much lower than the Alfv\'en velocity.

In the impulsive phase of this flare, magnetic energy release appears not in the form of a
quasi-steady reconnection annihilating anti-parallel magnetic field and thus producing outward
moving FPs. The main flare energy release at a given position in the arcade seems to last only a
short time (order of a few seconds) and moves along the arcade in a systematic manner.
The observed modulation of the HXR flux and the related anti-correlation of the spectral index in
each Elementary Flare Burst appear to be caused by spatial variations of the acceleration efficiency.
The temporal variations thus seem to be the result of a continuously moving trigger propagating
through variable conditions in the arcade. The short lifetime of a FPs at a given position shows that
particle trapping is not effective over timescales larger than several tens of seconds.

The observed simple and systematic motions set this event apart as a prototype for a type of HXR flare
evolving \emph{along the arcade}. The FP motions of this flare contradict clearly the expectations of
the standard 2D reconnection model. The fact that we do not observe a systematic increase
(up to the instrumental limits) of the
separation of the FPs, does challenge the idea that the reconnection points move upwards and particles
are accelerated in field lines successively farther out during the main HXR emitting phase of the
flare. A possible interpretation is that the trigger releases the main energy stored in a
two-dimensional loop structure within seconds, without noticeable FP motion, and moves on.
Reconnection in the given structure may still continue, but with HXR emission below RHESSI
sensitivity and at a much reduced energy release rate. Such secondary reconnection may be the
cause of decimetric radio emissions continuing for 6 minutes after 13:22 UT, the end of HXR
emission, and may produce the expansion of the two H$\alpha$ ribbons as observed in other flares.

We thus propose a scenario in which a disturbance, probably connected with the eruption of a filament,
propagates along the arcade like a burning fuse, sequentially triggering
reconnection and particle acceleration in the flare loops.
The main HXR emission from the FP reflects the propagation of this disturbance, not the
reconnection process at a given place in the arcade. If the dominating emission is strong
and short-lived, the local conditions cause the observed temporal modulation.

The global evolution may be compatible with the standard model of an eruptive flare, if one
allows the filament to erupt in such a way that one of its ends does not move while
the other starts to rise. In this scenario the reconnection process spreads along the arcade
until it reaches the end. The arcade erupts similar to the opening of a zipper, where the
lower side run across the arcade and the upper side is the filament.  Future studies of HXR FPs
in a large number of flares may establish such a scenario and stimulate the development
of 3D reconnection models needed to understand these observations.

\acknowledgements
The analysis of RHESSI data at ETH Zurich is partially supported by the Swiss
National Science Foundation (grant nr. 200020-105366). We thank the many people
who have contributed to the successful operation of RHESSI and acknowledge
S. Krucker for helpful discussions.

\end{document}